\documentclass[12pt]{article}
\usepackage{epsfig} 
\def\eqn#1{eq.~(\ref{#1})}

\newcommand{\be}{\begin{equation}} 
\newcommand{\ee}{\end{equation}} 
\newcommand{\OO}{{\cal O}} 
\newcommand{\as}{\alpha_s} 
 
\newcommand{\aoverpi}{\frac{\alpha_s}{2\pi}} 
\newcommand{\la}{\langle} 
\newcommand{\ra}{\rangle}

\begin{document}  

\begin{titlepage}
\begin{flushright}
IPPP/02/13 \\
DCPT/02/26
\end{flushright}
\vspace{0.5cm}
\begin{center}
{\Large{\bf Gluon induced contributions to $WZ$}} \\
\vskip 0.2cm
{\Large{\bf and $W\gamma$ production at NNLO} }
 
\vskip 1.2cm
{\bf K. L. Adamson$^{a}$, 
D. de Florian$^{b}$
and  A. Signer$^a$}
\vskip .8cm
\vskip .3cm
{\it $^a$ IPPP, Department of Physics, University of Durham,
Durham DH1 3LE, England}
\vskip .3cm
{\it $^b$ Departamento de F\'\i sica, Universidad de Buenos Aires, Argentina}
\vskip 4.cm

\vspace*{0.3cm}  

\hspace{1cm}
 
\large
{\bf Abstract} \\
\end{center}

We calculate the contribution of the partonic processes $gg
\rightarrow WZ q\bar{q}$ and $gg \rightarrow W\gamma q\bar{q}$ to $WZ$
and $W\gamma$ pair production at hadron colliders, including anomalous
triple gauge-boson couplings. We use the helicity method and include
the decay of the $W$ and $Z$-boson into leptons in the narrow-width
approximation. In order to integrate over the $q\bar{q}$ final state
phase space we use an extended version of the subtraction method to
NNLO and remove collinear singularities explicitly. Due to the large
gluon density at low $x$, the gluon induced terms of vector-boson pair
production are expected to be the dominant NNLO QCD correction,
relevant at LHC energies. However, we show that due to a cancellation
they turn out to provide a rather small contribution, anticipating
good stability for the perturbative expansion.

\vspace{0.5cm}
\normalsize
   \end{titlepage}
\newpage

\setcounter{page}{1}  
\pagestyle{plain}


\section{Introduction}  
  
A direct test of the non-abelian nature of the weak interactions  
can be made through the study of vector-boson pair production. These  
processes already involve triple gauge-boson couplings at tree level  
and, therefore, offer a promising framework for their study. In the  
Standard Model, the triple gauge-boson couplings are completely fixed  
by gauge symmetry. However, these couplings can be modified by new  
physics occurring at a higher energy scale. Given that the triple  
gauge-boson couplings have been much less precisely measured than,  
for example, the couplings of gauge bosons to fermions, it is natural to  
search for new physics by looking for anomalous triple gauge-boson  
couplings.   
  
Trilinear gauge-boson couplings have been studied at LEP2~\cite{lep2}  
and the Tevatron~\cite{tevatron}. It should be emphasized that the  
studies at hadron colliders are complementary to studies at  
electron-positron colliders. While bounds on anomalous couplings  
obtained from LEP2 data are more stringent than those from hadron  
colliders, the latter offer the possibility to study these couplings  
at a higher centre of mass energy, where effects of anomalous  
couplings are enhanced. The amount of available data from hadron  
colliders will dramatically increase with the Run~II at Tevatron and  
even more so with the LHC.  This will allow the study of triple  
gauge-boson couplings to be taken one step further. Of course, all possible  
vector-boson pair production processes will be analysed. However, in  
this paper we concentrate on $W^\pm\gamma$ and $W^\pm Z$ pair  
production.

In order to get reliable theoretical predictions for vector-boson pair  
production it is important to include higher-order corrections.  
One-loop QCD corrections, treating the vector bosons as stable  
particles, were computed for $W^\pm \gamma$ production  
\cite{stableWY} and $W^\pm Z$ production \cite{stableWZ} ten years  
ago. Since vector bosons are identified through their decay products  
it is essential to include the decay of the bosons into fermions. This  
allows arbitrary cuts to be added to the kinematics of the final states  
and therefore facilitates the comparison of theoretical predictions  
with experimental measurements. The calculation of these processes in  
the narrow-width approximation but retaining spin information via  
decay-angle correlations only in the real part of the amplitudes has been  
published in \cite{spinVV}. Anomalous couplings for the  
$W^\pm\gamma$~\cite{anomWY} and $W^\pm Z$~\cite{anomWZ} processes have  
also been included. Finally, analytic results of the one-loop amplitudes  
for vector-boson pair production with the subsequent decay into lepton  
pairs in the narrow-width approximation have been presented in  
\cite{dks1}. In \cite{dks2,dFs} these amplitudes have been used to  
obtain numerical results including all decay-angle correlations. Steps  
beyond the inclusion of $\OO(\as)$ corrections in the narrow-width  
approximation have been made in \cite{ce}, where non-doubly resonant  
diagrams have been included and in \cite{adp}, where one-loop  
logarithmic electroweak corrections have been calculated.  
  
The size of the one-loop QCD corrections depends crucially on the  
physical quantity under investigation. Usually they are of the order  
of a few 10\%, but in some particularly interesting cases, for example 
the high $p_T$ tail of transverse momentum distributions they can  
increase the leading order cross section by several 100\%.  The reason  
for such huge corrections is that the next-to-leading order  
correction to vector-boson pair production consists of two  
parts. First, there is the one-loop correction to the leading order  
partonic process $q\bar{q}\to VV$. Second, there are new partonic  
channels, $qg\to VVq$, that contribute to the cross section at NLO.  
Even though these processes are suppressed with respect to the leading  
order partonic process $q\bar{q}\to VV$ by the strong coupling  
constant $\as$, they can easily be more important numerically, since  
they are enhanced by the gluonic parton distribution function.  This  
is particularly the case for the LHC (or even a much higher energy 
Very Large Hadron Collider, or VLHC), where gluon  
induced processes become increasingly dominant.  Unfortunately, the  
kinematical regimes where these contributions are largest are exactly  
those where effects from anomalous couplings are expected to manifest  
themselves.  Therefore, it is important to gain a better control of  
the theoretical predictions in these kinematical regimes.  
  
One way to improve the situation is to impose a jet veto. This  
suppresses the effect of the new partonic channels that are opened at  
higher order \cite{anomWZ,dks2,dFs}. On the other hand, such a veto  
also reduces the amount of data considerably. Another possibility is  
to include even higher order terms. Of course, a full NNLO calculation of  
vector-boson pair production is not to be expected in the  
near future. Such a calculation would have to include two-loop  
corrections to the partonic subprocesses $q\bar{q}\to VV$, one-loop  
corrections to $qg\to VVq$ and, finally, (amongst others) 
processes with $gg$ in the  
initial state. Motivated by the observation that loop corrections to  
partonic processes tend to be of the expected moderate size and huge  
corrections are mainly due to the opening of new channels, in this  
paper we include all NNLO terms that are maximally enhanced by the  
gluon distribution functions.  
  
In general, there are two classes of such contributions. First, there  
is the process $gg\to VV$. These processes have been studied
previously~\cite{GlVdB}. The amplitude for this process vanishes at  
tree level but may be non-zero at one-loop. In the case of  
$W^\pm\gamma$ and $W^\pm Z$ pair production, however, the amplitude  
for this process vanishes at all orders, due to charge  
conservation. Second, there are the processes $gg\to VVq\bar{q}$.  The  
amplitudes for these processes have been calculated in \cite{VVqqampl}  
and have been used to compute cross sections for the production of a  
vector boson pair together with two jets. We extend this work by not  
requiring any jets to be observed in the final state. Hence, the  
amplitudes $gg\to VVq\bar{q}$ have to be integrated over the  
$q\bar{q}$ final state phase space. This results in infrared  
singularities that will have to be absorbed in the parton distribution  
functions. Furthermore, we also extend the previous work by adding  
anomalous couplings.  
  
We find that the contribution of the gluon induced processes to $WZ$ and 
$W\gamma$ production is smaller than may have been anticipated, due to 
a sign change in the hard scattering part.
We also find that this NNLO term is affected less by anomalous couplings than
the corresponding LO and NLO terms.

\section{Calculation}  

Vector boson pair production has been calculated up to NLO in the
helicity method.  However, in an NLO calculation for $pp \rightarrow
WZ$, whilst the process $\bar{q}q \rightarrow WZ$ is included at NLO,
the process $qg \rightarrow WZq$ is only calculated at LO.  This means
that we only have a leading order calculation in the circumstances
where gluon-induced subprocesses are important, as they are expected
to be at the LHC due to the large gluon density at low $x$.  Therefore
we want to include the next order in $\as$, the process $gg
\rightarrow WZq\bar{q}$ (and similarly for $W\gamma$).

For $W^-Z$ production we will use the labelling $\, g_1 \, g_2 \, l_3
\, \bar{\nu}_4 \, \bar{l}'_5 \, l'_6 \, q_7 \, \bar{q}_8$, where $g_1$
and $g_2$ are the incoming gluons, leptons 3 and 4 are the decay
products of the $W$ and 5 and 6 are from the $Z$. Clearly, $W\gamma$ will
have one less particle but this is a trivial modification.  It is also
straightforward to transform the calculation for other vector bosons.
  
In calculating $gg \rightarrow WZ$, we use a version of the  
subtraction method \cite{fks} to cancel the infrared singularities  
analytically and evaluate the finite remainders numerically.  We 
show the finite parts of the cross section in 
equations~(\ref{eq:gg8}), (\ref{eq:gg7}) and (\ref{eq:gg6}).  
  
A generic differential cross section is written as:  
\be  
d\sigma_{AB}(p_A,p_B)=\sum_{ab}\int d x_1 \, d x_2 \,f_{a/A} (x_A)  
f_{b/B} (x_B) \, d \hat{\sigma}_{ab}(x_A p_A,x_B p_B) \, ,  
\ee  
where $A$ and $B$ are hadrons while $a$ and $b$ are partons and  
$f_{a/A}$ and $f_{b/B}$ are the parton distribution functions.  The  
subtracted partonic cross section, $d\hat{\sigma}$, is finite for any  
infrared safe observable.  
  
In a general case, in order to obtain $d\hat{\sigma}$, soft and final
state collinear singularities have to be cancelled with the
singularities coming from virtual corrections. The remaining initial
state collinear singularities have to be factored out into the parton
distribution functions.  In an extension of the subtraction method to
NNLO, this demands the subtraction of all possible singularities as
well as their analytical integration in the corresponding limits. Up
to now, such a general framework has not been developed. Only a
limited number of analytical calculations to NNLO accuracy have been
performed.

The main difficulty for the implementation of a subtraction procedure
at NNLO is the appearance of several new (multiple parton) kinematical
configurations where soft and/or collinear singularities arise.  These
are: the triple collinear case (three partons become collinear), the
double soft case (two partons are soft), the soft-collinear case (one
parton is soft and the other collinear) and the double collinear case
(two pairs of partons are independently collinear).  The single soft
and collinear singularities, which appear also at NLO, have to be
added to that list.

Even though the behaviour of tree and one-loop amplitudes in those
limits has been recently obtained \cite{nnlolimits}, it is not yet
clear how to unify all singularities in a single subtraction term, or
how to subtract all of them independently while avoiding double
counting and allowing the analytical integration of the subtracted
terms to be performed.
 
Our case, however, is particularly simple. Since we include only $gg$
induced processes at NNLO there are no final state collinear
singularities (see below about final state singularities in the
$W\gamma$ case). There are also no soft singularities, since there are
no gluons (or $q\bar{q}$ pairs) in the final state.  Therefore, the
only singularities that arise are the single collinear ones, which can
be treated in a similar way as a typical NLO calculation, and the {\rm
double collinear} ones, i.e., the collinear splitting of {\rm each}
initial state gluon into a $q\bar{q}$ pair, a `genuine' NNLO
contribution.
 
In order to absorb the  remaining initial state collinear
singularities we use    
\be  
\label{eq:pd}  
f_{a/d}=\delta_{ad} \delta (1-x)-\frac{\alpha_s}{2\pi}\frac{1}  
{\bar{\epsilon}} P_{ad}(x,0) -   
\left(\frac{\alpha_s}{2\pi}\right)^2 Q_{ad}(x) \, . 
\ee    
This is the NNLO expression in $\overline{\textrm{MS}}$ and as usual
we define $\frac{1} {\bar{\epsilon}} \equiv \frac{1}{\epsilon} -
\gamma_E + \log 4\pi$.  $P_{ad}(x,0)$ is the one-loop Altarelli-Parisi
splitting function for $\varepsilon =0$ (4 dimensions). The ${\cal
O}(\alpha_s^2)$ term $Q_{ad}$, including contributions from the
two-loop splitting functions,
 does not actually contribute in our case.  
  
Comparing subtracted and unsubtracted terms and using \eqn{eq:pd},  
we obtain the relation:    
\begin{eqnarray}  
\lefteqn{d\hat{\sigma}_{gg}^{(2)}(k_1,k_2)=d\sigma_{gg}^{(2)}(k_1,k_2)}  
&& \nonumber\\  
&+&\aoverpi\int d x_1 \left(\frac{1}{\bar{\epsilon}}P_{qg}(x_1)\right)   
d\sigma_{qg}^{(1)}(x_1 k_1,k_2)  
+\aoverpi \int d x_1 \left(\frac{1}{\bar{\epsilon}}P_{\bar{q}g}(x_1)  
\right) d\sigma_{\bar{q}g}^{(1)}(x_1 k_1,k_2) \nonumber \\  
&+&\aoverpi \int d x_2\left(\frac{1}{\bar{\epsilon}}P_{qg}(x_2)\right)  
 d\sigma_{gq}^{(1)}( k_1,x_2 k_2)  
+ \aoverpi \int d x_2\left(\frac{1}{\bar{\epsilon}}P_{\bar{q}g}(x_2)  
\right)d\sigma_{g\bar{q}}^{(1)}( k_1,x_2 k_2) \nonumber \\  
&+&\left(\aoverpi\right)^2  \int d x_1  
d x_2  \left(\frac{1}{\bar{\epsilon}}P_{qg}(x_1)\right)\left(\frac{1}  
{\bar{\epsilon}}P_{\bar{q}g}(x_2)\right) d\sigma_{q\bar{q}}^{(0)}  
(x_1 k_1,x_2 k_2)  \nonumber\\  
&+&\left(\aoverpi\right)^2  \int d x_1  
d x_2 \left (\frac{1}{\bar{\epsilon}}P_{\bar{q}g}(x_1)\right)\left(\frac{1}  
{\bar{\epsilon}}P_{qg}(x_2)\right) d\sigma_{\bar{q}q}^{(0)}(x_1 k_1,x_2 k_2)  
\label{sigma_hat_1} \, , 
\end{eqnarray}  
where $k_1,k_2$ are the momenta of the incoming gluons  1 and 2.  
The superscript (2) denotes the NNLO term, while (1) is NLO and (0) is   
leading order. In \eqn{sigma_hat_1} all terms on the right hand side  
are separately divergent but the sum is finite.  
  
Once we have this expression, we can write out the counterterms   
in terms of energy and angle variables and cancel the poles explicitly.  
Following \cite{fks} we use the parameterisation  
\begin{eqnarray}  
k_1&=&\frac{\sqrt{s_{12}}}{2}(1,\vec{0},1) \\  
k_2&=&\frac{\sqrt{s_{12}}}{2}(1,\vec{0},-1) \\  
k_i&=&\frac{\sqrt{s_{12}}}{2}\xi_i\left(1,\sqrt{1-y_i^2}\vec{e}_T,y_i\right)
\, ,  
\end{eqnarray}  
where $k_i$ is the momentum of an outgoing (anti)quark,  
$\vec{e}_T$ is a unit vector in transverse momentum space, $y_i=\cos  
\theta_i$ is the angle variable  $-1\le y_i \le 1$ and  $\xi_i$ is the  
rescaled energy variable $0 \le \xi_i \le 1$. Performing part of the  
phase-space integration analytically we can rewrite \eqn{sigma_hat_1}  
as a sum of three separately finite pieces  
\begin{equation}  
d\hat{\sigma}_{gg}^{(2)}=d\sigma^{({\rm fin},8)}+  
d\sigma^{({\rm fin},7)}+d\sigma^{({\rm fin},6)} \, , 
\end{equation}  
with terms $d\sigma^{({\rm fin},8,7,6)}$ denoting the 8,7 and 6-parton finite   
parts. The explicit form of  $d\sigma^{({\rm fin},8)}$ is given by  
\begin{eqnarray}  
d\sigma^{({\rm fin},8)} & = &   
(1-y_7^2)(1-y_8^2){\cal{M}}_{gg}^{(8)}\left(k_1,k_2,  
\{k_i\}_{3,6},k_7,k_8\right)\frac{s_{12}^2}{64(2\pi)^6}    
\xi_7 \, \xi_8 \, {\cal{P}}(y_7) \, {\cal{P}}(y_8)  \nonumber      
\\ & & d\xi_7 \,  
d\xi_8  \, d y_7 \, d y_8 \, d\varphi_7 \, d\varphi_8 \,   
d\Phi_{(3-6)} \label{eq:gg8}  \, , 
\end{eqnarray}  
where $d\Phi_{(3-6)}$ denotes the phase-space integration over the  
vector-boson decay products. ${\cal{M}}_{gg}^{(8)}$ denotes the  
squared amplitude summed (averaged) over helicities, including the  
flux factor $1/(2 s_{12})$.  ${\cal{P}}(y_i)$ is given by:
\begin{equation}  
{\cal{P}}(y_i)\equiv\frac{1}{2}\Big[ \Big(\frac{1}{1-y_i}\Big)_{\delta_I}+  
\Big(\frac{1}{1+y_i}\Big)_{\delta_I} \Big]  \,. 
\end{equation}  
The distributions $\left(\frac{1}{1\pm y_i}\right)_{\delta_I}$ have  
been introduced in \cite{fks} and are defined for an arbitrary test  
function $f(y_i)$ through  
\be  
\label{distribution_def}  
\langle \left(\frac{1}{1\pm y_i}\right)_{\delta_I}, f(y_i) \rangle  
= \int_{-1}^{1} dy_i \,   
\frac{f(y_i) - f(\mp1) \theta(\mp y_i-1+\delta_I)}{1\pm y_i} \,. 
\ee  
The two ${\cal{P}}$ distributions in eq.(\ref{eq:gg8}) perform the subtraction of  
both double and single collinear singularities of NNLO amplitudes, leaving a  
(numerically) integrable remnant. 
 
The 7-parton finite part reads  
\begin{eqnarray}  
d\sigma^{({\rm fin},7)} & = & \aoverpi \left({\cal{L}}_7 P_{qg}^< (1-\xi_7) -   
P_{qg}'^< (1-\xi_7)\right) (1-y_8^2) \nonumber\\  
& &  \Big\{{\cal{M}}_{\bar{q}g}^{(7)} \left((1-\xi_7)k_1,k_2,  
\{k_i\}_{3,6},k_8\right)+{\cal{M}}_{g\bar{q}}^{(7)} \left(k_1,(1-\xi_7)k_2,  
\{k_i\}_{3,6},k_8\right)\Big\} \nonumber\\  
& & \frac{s_{12}}{8 (2\pi)^3} \, \xi_8 \, {\cal{P}}(y_8) \, d\xi_7 \,   
d\xi_8 \, d y_8 \, d\varphi_8 \, d\Phi_{(3-6)} \nonumber\\  
&  & + \, \aoverpi \left({\cal{L}}_8 P_{\bar{q}g}^< (1-\xi_8) -   
P_{\bar{q}g}'^< (1-\xi_8)\right) (1-y_7^2) \nonumber\\  
& &  \Big\{{\cal{M}}_{qg}^{(7)} \left((1-\xi_8)k_1,k_2,  
\{k_i\}_{3,6},k_7\right)+{\cal{M}}_{gq}^{(7)} \left(k_1,(1-\xi_8)k_2,  
\{k_i\}_{3,6},k_7\right)\} \nonumber\\  
& & \frac{s_{12}}{8 (2\pi)^3} \, \xi_7 \, {\cal{P}}(y_7) \, d\xi_7 \,   
d\xi_8 \, d y_7 \, d\varphi_7 \, d\Phi_{(3-6)} \, , 
\label{eq:gg7}  
\end{eqnarray}  
where we introduced   
\begin{equation}  
{\cal{L}}_7=\log \frac{s_{12} \, \delta_I \, \xi_7^2}{2 \mu^2} \qquad  
{\cal{L}}_8=\log \frac{s_{12} \, \delta_I \, \xi_8^2}{2 \mu^2} \, ,   
\end{equation}  
and split up the unregularized Altarelli-Parisi function into the  
4-dimensional piece, $P_{ad}^<$, and the piece proportional to  
$\varepsilon$,  $P_{ad}'^<$. For completeness we give the explicit  
form   
\begin{eqnarray}  
 P_{qg}^< (1-\xi_i)&=&\frac{1}{2}(1-2 \xi_i +2 \xi_i^2) \\  
P_{qg}'^<(1-\xi_i) &=& \xi_i(\xi_i-1) \,. 
\end{eqnarray}  
In this case, the single collinear singularities of the NLO amplitudes are  
subtracted by the appearance of a single ${\cal{P}}$ distribution. 
  
Finally, the 6-parton finite piece is  
given by  
\begin{eqnarray}  
d\sigma^{({\rm fin},6)} & = & \left(\aoverpi\right)^2 \,   
\left({\cal{L}}_7 P_{qg}^< (1-\xi_7)-P_{qg}'^<(1-\xi_7)\right) \,   
\left({\cal{L}}_8 P_{\bar{q}g}^<  
(1-\xi_8)-P_{\bar{q}g}'^<(1-\xi_8)\right) \nonumber\\   
& & \Big\{{\cal{M}}_{\bar{q}q}^{(6)}\left((1-\xi_7)k_1,(1-\xi_8)k_2,  
\{k_i\}_{(3,6)}\right) \nonumber\\  
& &\quad + \, {\cal{M}}_{q\bar{q}}^{(6)}\left((1-\xi_8)k_1,(1-\xi_7)k_2,  
\{k_i\}_{(3,6)}\right)\Big\}  
\, d\xi_7 \, d\xi_8  \, d\Phi_{(3-6)} \,. 
\label{eq:gg6}  
\end{eqnarray}  
  
We note that $\delta_I$ is an arbitrary quantity and even though  
$d\sigma^{({\rm fin},8)}, d\sigma^{({\rm fin},7)}$ and $d\sigma^{({\rm  
fin},6)}$ depend on it, this dependence cancels exactly in  
$d\hat{\sigma}_{gg}^{(2)}$. Hence, making sure that the numerical  
results are independent of $\delta_I$ is a useful check for our  
Monte Carlo program.  
  
In calculating the cross section as above, we use 6 parton and 7  
parton amplitudes as given in \cite{dks1}.  We also require the 8 and  
7 parton amplitudes for $gg \rightarrow q\bar{q}WZ$ and $gg  
\rightarrow q\bar{q}W\gamma$ respectively. These amplitudes have been  
calculated previously~\cite{VVqqampl}.  We  recalculated these amplitudes  
using the helicity method and including anomalous couplings.  
For the triple gauge vertex for $W^+_\alpha(p_+)W^-(p_-)_\beta   
V_\mu(q)$, including the anomalous terms, we use   
\begin{eqnarray}  
\Gamma^{\alpha \beta \mu}_{WWV}(p_+,p_-,q)& = & \Big((p_+-q)^\alpha   
g^{\beta\mu}\frac{1}{2}(g_1^V+\kappa^V+\lambda^V\frac{p_-^2}{M_W^2})  
\nonumber \\ & &   
+(q-p_-)^\beta g^{\alpha \mu}\frac{1}{2}(g_1^V+\kappa^V +\lambda^V  
\frac{p_+^2}{M_W^2}) \nonumber \\& &   
+(p_+-p_-)^\mu(-g^{\alpha\beta}\frac{1}{2} q^2 \frac{\lambda^V}  
{M_W^2}+\frac{\lambda^V}{M_W^2}q^\alpha q^\beta)\Big),  
\end{eqnarray}  
where $V$ can be $Z$ or $\gamma$ and all momenta are taken to be  
outgoing.  In the Standard Model, $g_1^V=\kappa^V=1$ and  
$\lambda^V=0$. In order to preserve electromagnetic gauge invariance  
we always set $g_1^\gamma$ to the Standard Model value 1.  

We also make use of form factors to avoid the violation of unitarity
that would otherwise result from this vertex at high energies.  The
form factors used is are the conventional ones:
\begin{equation}
\Delta g_1^V \rightarrow \frac{\Delta g_1^V}{(1+\hat{s}/\Lambda^2)^2}, 
\quad
\Delta \kappa^V \rightarrow \frac{\Delta \kappa^V}{(1+\hat{s}/\Lambda^2)^2}, 
\quad
\Delta \lambda^V \rightarrow 
\frac{\Delta \lambda^V}{(1+\hat{s}/\Lambda^2)^2}
\label{formfactor}
\end{equation}
where $\Lambda$ is the scale of the new physics that causes the anomalous 
couplings.

Photons with large transverse momentum can be produced in hadronic
collisions not only directly, but also, and significantly, from the
fragmentation of a final state parton. A full perturbative calculation
of a process involving the production of photons should in principle
include the calculation of both {\rm direct} and {\rm fragmentation}
components, since only their sum is physically well defined beyond
LO. A calculation of the fragmentation part is not even available at
NLO for the process of interest in this paper.  Fortunately there is a
way to suppress the fragmentation contribution, which actually
constitutes a background to the search of anomalous couplings, by
requiring the photons to be isolated from the hadrons.  The usual way
to perform the isolation is to require the transverse hadronic
momentum in a cone around the photon to be smaller than a fraction of
the transverse momentum of the photon. In this way, the contribution
of the fragmentation component can be reduced to the percent level.
In this paper, we will use the isolation procedure introduced by
Frixione \cite{Frixione:1998jh} which allows us to {\rm completely}
suppress the fragmentation component.  Therefore, we reject all events
unless the transverse hadronic momentum deposited in a cone of size
$R_0$ around the momentum of the photon fulfills the following
condition
\begin{equation} 
\label{eq:isolation} 
\sum_i p_{Ti}\, \theta(R -R_{i\gamma}) \le p_{T\gamma}  
\left( \frac{1-\cos R}{1-\cos R_0}\right) \,, 
\end{equation} 
for all $R \le R_0$, where the `distance' in pseudorapidity and azimuthal 
 angle is defined by $R_{i\gamma}=\sqrt{(\eta_i-\eta_\gamma)^2 +  
(\phi_i-\phi_\gamma)^2 }$. 
In this way, only soft partons can be emitted collinearly to the
photon in the direct contribution and, therefore, no final state
quark-photon collinear singularity arises. For the purposes of our
computation this allows us to perform the calculation of the $gg$
initiated contribution at NNLO without needing to do any subtraction
of the corresponding singularity. While the possibility of performing
such isolation at the experimental level is under investigation, the
choice of this particular method will not alter at all the conclusions
of our calculation in the $W\gamma$ channel.  Particularly, we have
checked that the results at NLO using the procedure in
eq.(\ref{eq:isolation}) are very close to the ones obtained performing
the usual `cone' isolation procedure \cite{dFs}.

In this calculation, contributions from $b$ and $t$ quarks have been
neglected.  It is assumed that these will be suppressed by the large
top quark mass. This might not be a particularly good approximation
for large energies, but in view of the smallness of the gluon induced
corrections a more detailed treatment doesn't seem to be justified.

\section{Numerical Results}  
  
For the numerical results presented in this section we use the
MRST~2001 parton distribution functions~\cite{mrst} with the one-loop
expression for the coupling constant ($\alpha_s(M_Z) = 0.119$).  The
factorization and renormalization scales are fixed to
$\mu_F=\sqrt{\hat{s}}$, $\mu_R=\sqrt{\frac{1}{2}(M_W^2+M_Z^2) +
\frac{1}{2}(p_{T\,W}^2+p_{T\,Z}^2)}$ in the case of $WZ$ production
and to $\mu_F=\mu_R= \sqrt{M_W^2 + p_{T\, \gamma}^2} $ in the
$W\gamma$ case.  Notice that in order to compute the $gg$ induced
contribution we also use the same NLO combination of parton
distribution function and coupling constant.  The effect of changing
to NNLO parton distributions, not fully available yet, is expected to
be small and will not alter the conclusion of our analysis.

The masses of the vector bosons have been set to $M_Z=91.187$~GeV and
$M_W=80.41$~GeV. We do not include any electroweak corrections, but
choose the coupling constants $\alpha$ and $\sin^2\theta_W$ in the
spirit of the improved Born approximation~\cite{impr_born}. For the
couplings of the vector bosons with the quarks we use $\alpha =
\alpha(M_Z)= 1/128$ whereas for the photon coupling we use $\alpha =
1/137$. We neglect contributions from initial state top quarks and use
the following values for the Cabibbo-Kobayashi-Maskawa matrix
elements: $|V_{ud}|=|V_{cs}|=0.975$ and
$|V_{us}|=|V_{cd}|=0.222$. Note that we do not include the branching
ratios for the decay of the vector bosons into leptons.

For all plots we use a set of standard cuts. For charged leptons we  
require $p_T>20$~GeV and $\eta < 2.5$. In addition, we require a  
missing transverse momentum $p_T^{\rm miss}>20$~GeV. The photon  
transverse momentum cut we use is $p_T^\gamma > 20$~GeV,  
while for the isolation prescription in \eqn{eq:isolation} we set 
$R_0=1$.

\begin{figure}[htb]
\begin{center}
\begin{tabular}{c}
\epsfxsize=8.5truecm
\epsffile{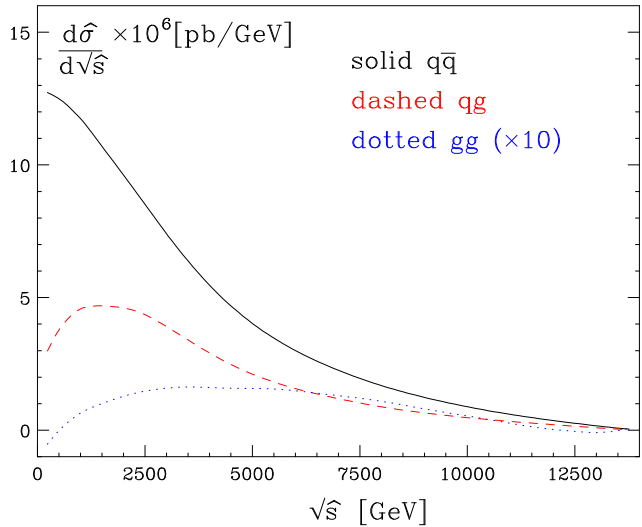}\\
\end{tabular}
\end{center}
\caption{\label{figsWY}{\em Partonic cross sections for $W\gamma$:
$\hat{\sigma}_{q\bar{q}}$, $\hat{\sigma}_{qg}$ and
$\hat{\sigma}_{gg}$.  }}
\end{figure}

\begin{figure}[htb]
\begin{center}
\begin{tabular}{c}
\epsfxsize=8.5truecm
\epsffile{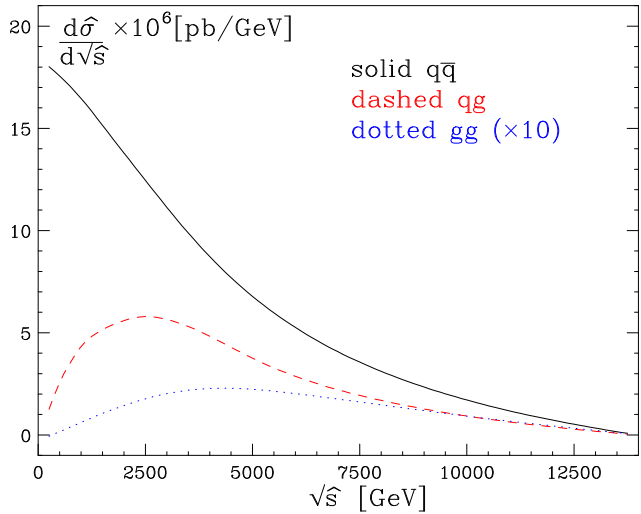}\\
\end{tabular}
\end{center}
\caption{\label{figsWZ}{\em Partonic cross sections for $WZ$:
$\hat{\sigma}_{q\bar{q}}$, $\hat{\sigma}_{qg}$ and $\hat{\sigma}_{gg}$   }}
\end{figure}

In Figures~\ref{figsWY} and~\ref{figsWZ} we show the $\overline{MS}$
subtracted partonic cross sections for the different initial states
$q\bar{q},\, qg$ and $gg$. These are produced by binning the results
at 14 TeV by $\sqrt{\hat{s}}$.  It can be seen that the outcomes for
$W\gamma$ and $WZ$ are very similar. For better visibility
$\hat{\sigma}_{gg}$ has been increased by a factor 10 in the plot. 

As can be observed, there is a reasonably good perturbative
convergence at the level of the partonic cross section, considering
that $\hat{\sigma}_{q\bar{q}}$ is of ${\cal O}(1+\alpha_s)$,
$\hat{\sigma}_{qg}$ of ${\cal O}(\alpha_s)$, and $\hat{\sigma}_{gg}$
of ${\cal O}(\alpha_s^2)$, with $\alpha_s\sim 0.1$ for the typical
scales of these processes.  In the $\overline{MS}$ both the $q\bar{q}$
and $qg$ contributions are positive in the whole range of $\hat{s}$,
while the new piece, the $gg$ partonic cross section becomes negative
only for small values of $\hat{s}$, close to the threshold for the
production of the vector bosons.

\begin{figure}[h]
\begin{center}
\begin{tabular}{c}
\epsfxsize=8.5truecm
\epsffile{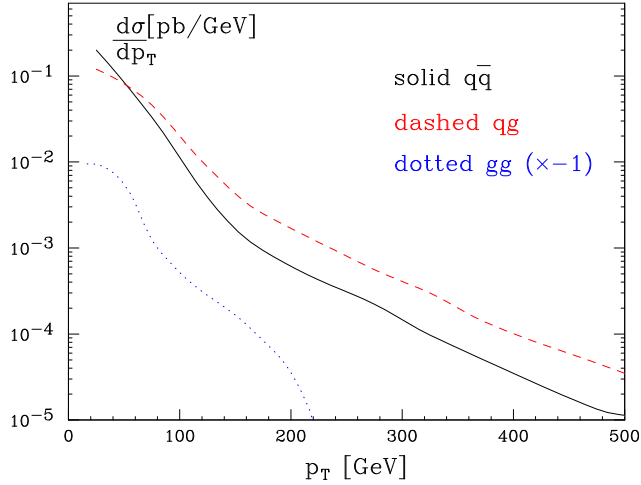}\\
\end{tabular}
\end{center}
\caption{\label{figptWY}{\em $W\gamma$ production: $p_T$ distribution
for LHC, $q\bar{q}, qg$ and $gg$ pieces separately  }}
\end{figure}

\begin{figure}[h]
\begin{center}
\begin{tabular}{c}
\epsfxsize=8.5truecm
\epsffile{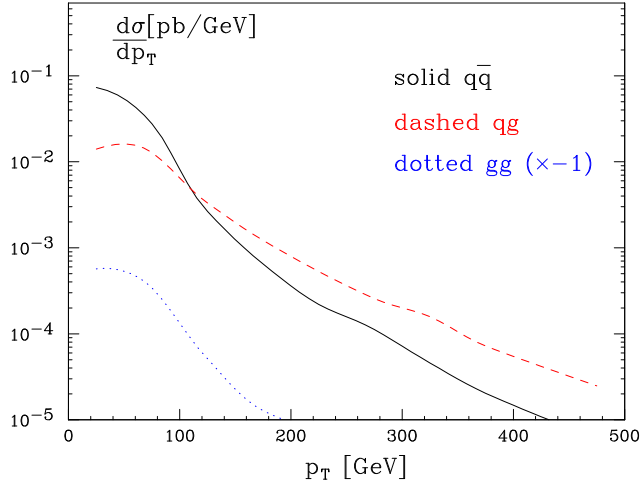}\\
\end{tabular}
\end{center}
\caption{\label{figptWZ}{\em $WZ$ production: $p_T$ distribution for
LHC, $q\bar{q}, qg$ and $gg$ pieces separately }}
\end{figure}

The situation changes considerably when the physical hadronic cross
section is computed, shown in Figures~\ref{figptWY} and~\ref{figptWZ}.
In Figure~\ref{figptWY} we plot the transverse momentum distribution
of the photon, whereas in Figure~\ref{figptWZ} we see the $p_T$ of the
lepton produced by the decay of the $W$.  Again the form of the
distributions is very similar for the $WZ$ as compared to the
$W\gamma$ case.
  
As indicated in the introduction, because of the large $qg$
luminosity, the hadronic contribution due to this initial state
becomes even larger than the `leading' $q\bar{q}$ contribution. This
is particularly clear at large transverse momentum, where the relative
increase in the luminosity exceeds the effect of the suppression due
to the extra $\alpha_s$ coupling observed at the partonic level.  This
is not the case for the $gg$ contribution, which remains rather small,
and for the observable studied here provides a negative contribution
to the hadronic cross section.

The features of both $qg$ and $gg$ contributions to $WZ$ and $W\gamma$
production can be qualitatively understood in the following way: the
$qg$ and $gg$ luminosities are steep functions of the momentum
fraction carried by the partons, mostly due to the fast increase of
the gluon distribution $g(x)$ when $x$ decreases. Even though the
hadronic center of mass energy $S$ is very large, the average value of
the partonic one $\la\hat{s}\ra =\la x_1 x_2\ra S$ can be considerably
smaller, and actually closer to the minimum needed to produce the
gauge bosons with the required transverse momentum. Therefore in the
convolution between the parton distributions and the partonic cross
sections shown in Figures~\ref{figsWY} and~\ref{figsWZ}, the hadronic
result will mostly pick up the features of the partonic cross section
at low values of $\hat{s}$. In the $qg$ case, the partonic cross
section is positive and has its maximum at low $\hat{s}$, giving as a
result a large hadronic contribution. In the $gg$ case, the partonic
cross section has a change of sign at low values of $\hat{s}$, with
the corresponding compensation between negative and positive
contributions when convoluting with the parton densities.  Therefore
the $gg$ hadronic contribution results in a negative and small
correction to the NLO cross section.
 
\begin{figure}[htb]
\begin{center}
\begin{tabular}{c}
\epsfxsize=8.5truecm
\epsffile{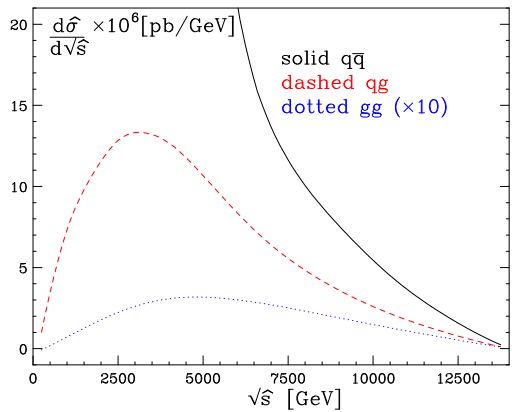}\\
\end{tabular}
\end{center}
\caption{\label{figsAC}{\em Partonic cross section
$\hat{\sigma}_{q\bar{q}}$, $\hat{\sigma}_{qg}$ and $\hat{\sigma}_{gg}$
for $WZ$ production with anomalous couplings as given in
\eqn{anom_couplings}  }} 
\end{figure}

In order to investigate whether the inclusion of anomalous couplings
changes this picture substantially, we show in Figure~\ref{figsAC} the
partonic cross sections with the following anomalous couplings:
\begin{equation}
g_1^\gamma=1; \quad g_1^Z=1.13; \quad \kappa^\gamma=1.2; \quad \kappa^Z=1.07;
\quad \lambda^\gamma=\lambda^Z=0.1
\label{anom_couplings}
\end{equation}
We use the form factors as given in \eqn{formfactor} and we take
$\Lambda$, the scale of new physics, to be 2~TeV. 
Figure~\ref{figsAC} is given with the same scale as
Figure~\ref{figsWZ} for comparison.  It is clear that, with the values
for the couplings that we have chosen, we do not see a significant
enhancement of the gluon induced term.  In fact, the gluon induced
corrections become even less relevant, as the $qg$ and especially the
$q\bar{q}$ terms are increased substantially.

In addition to these results, we also considered a hypothetical VLHC
or Very Large Hadron Collider, which would run at a centre of mass
energy of 200 TeV, to see whether the $gg$ part would become
significant at the higher energy (cuts etc. all remained the same as
previously).  It was found that the $gg$ part did not become more
important. The contributions from the $gg$ term remained at the 1\%
level.

\section{Conclusions}  

We have presented expressions for the finite cross section for
vector-boson pair production from a gluon-gluon initial state.  We
have then implemented these terms and the appropriate matrix elements
into a Monte Carlo integration.

We find the contribution of the gluon induced terms in $WZ$ and
$W\gamma$ production to be surprisingly small. Even the addition of
anomalous couplings enhances the $gg$ induced term substantially less
than the $q\bar{q}$ and $qg$ terms.  We have shown that this is due to
a change of sign in the hard scattering.

This is a good result from the point of view of experimental
predictions of vector-boson pair production, at least in the $WZ$ and
$W\gamma$ cases, as the impact of NNLO terms does not appear to be
particularly significant.  It also seems that the $gg$ term will not
be amplified excessively by anomalous couplings, though we have only
taken an example of anomalous couplings and have not made a detailed
study.

\subsection*{Acknowledgments}

This work was partly supported by the EU Fourth Framework Programme
`Training and Mobility of Researchers', Network `Quantum
Chromodynamics and the Deep Structure of Elementary Particles',
contract FMRX-CT98-0194 (DG 12 - MIHT) and by the EU Fifth Framework
Programme `Improving Human Potential', Research Training Network
`Particle Physics Phenomenology at High Energy Colliders', contract
HPRN-CT-2000-00149. KLA acknowledges support from a PPARC
studentship. DdF is supported by Conicet and Fundaci\'on Antorchas.
  


\end{document}